\documentstyle[aps,twocolumn]{revtex}

\begin{document}
\title{On the derivation of the $t-J$ model: electron spectrum and exchange
interactions in narrow energy bands}
\author{V.Yu.Irkhin$^*$}
\address{Institute of Metal Physics, 620219 Ekaterinburg, Russia}
\maketitle

\begin{abstract}
A derivation of the $t-J$ model of a highly-correlated solid is given
starting from the general many-electron Hamiltonian with account of the
non-orthogonality of atomic wave functions. Asymmetry of the Hubbard
subbands (i.e. of ``electron'' and ``hole''cases) for a nearly half-filled
bare band is demonstrated. The non-orthogonality corrections are shown to
lead to occurrence of indirect antiferromagnetic exchange interaction even
in the limit of the infinite on-site Coulomb repulsion. Consequences of this
treatment for the magnetism formation in narrow energy bands are discussed.
Peculiarities of the case of ``frustrated'' lattices, which contain
triangles of nearest neighbors, are considered.
\end{abstract}


The problem of strong correlations and magnetism in many-electron (ME)
systems is one of the most important in the solid state theory. Derivation
of the simplest ME model describing these phenomena in the case of narrow
energy bands (strong Coulomb interaction) was proposed by Hubbard\cite{Hub}.
This model was studied in a great number of papers.

A detailed investigation of ferromagnetism in the Hubbard model in the limit
of the infinite on-site Coulomb repulsion $U$ was performed by Nagaoka\cite
{Nag}. He proved rigorously that the ground state for the simple cubic and
bcc lattices in the nearest-neighbor approximation with the number of
electrons $N_e=N\pm 1$, $N$ being the number of lattice sites, possesses
maximum total spin, i.e. is saturated ferromagnetic, since this ordering
provides the maximum kinetic energy gain for an excess electron (hole). (The
same statement holds for the fcc and hcp lattices with the transfer integral 
$t<0$, $N_e=N+1$, or $t>0,N_e=N-1$.) The picture of saturated ferromagnetism
is preserved at small, but finite concentrations of current carriers $%
c=|N_e/N-1|$. In the case of a half-filled band ($N_e=N$), $|t|\ll U$ the
ground state is antiferromagnetic because of the Anderson's kinetic exchange
(superexchange) interaction\cite{And} which occurs in the second order in $%
|t|/U$. This interaction is due to the gain in the kinetic energy at virtual
transitions of an electron to a neighbor site, which are possible provided
that the electron at that site has an opposite spin direction. In systems
with a large finite $U$ and $N_e\neq N$, a competition between ferro- and
antiferromagnetic ordering occurs. Nagaoka\cite{Nag} has put forward a
criterion of ferromagnetism, which was based on the condition of the
spin-wave spectrum stability (the magnon spectrum can be calculated also
from the spin Green's function, see, e.g., Ref.\cite{IKJP85}). This
criterion has the form 
\begin{equation}
\alpha c>|t|/U  \label{Nf}
\end{equation}
where the constant $\alpha $ of the order of unity depends on the lattice
structure, $\alpha = 0.246$ for the simple cubic lattice. 
At the same time, pure antiferromagnetism is stable at $N_e=N$
only, and a phase separation\cite{Vish} takes place provided that the
condition (\ref{Nf}) is violated.

A version of the Hubbard model in the limit $U\rightarrow \infty $ with
inclusion of the antiferromagnetic exchange interaction $J$, the $t-J$
model, is widely applied now to describe copper-oxide high-$T_c$
superconductors and related systems (see the review\cite{tJ}). The quantity $%
J$ is usually related to the Anderson's exchange ($J=-2t^2/U$), but
sometimes it is considered as an independent phenomenological parameter.
Recently the difference between the $t-J$ model and large-$U$ expansion of
the Hubbard model has been discussed\cite{Esk}.

In his pioneering work\cite{Hub} Hubbard neglected the difference between
the atomic functions and orthogonal Wannier functions and did not treat the
non-orthogonality problem. At the same time, in the limit of large $U$ it is
more natural to use well-localized atomic wavefunctions. In the present
paper we develop a consistent formulation of a general ME model with strong
correlations with the use of the Bogoliubov approach\cite{Bog} and discuss
the electron spectrum picture and magnetism formation in this model.

We start from the general Hamiltonian of the ME system in a crystal 
\begin{equation}
{\cal H}=\sum_i\left( -\frac{\hbar ^2}{2m}\Delta _{{\bf r}_i}+V({\bf r}%
_i)\right) +\frac 12\sum_{i\neq j}\frac{e^2}{{\bf |r}_i{\bf -r}_j{\bf |}}
\label{HV}
\end{equation}
where 
\begin{equation}
V({\bf r})=-\sum_\nu \frac{Ze^2}{|{\bf r-R}_\nu |}=\sum_\nu v({\bf r-R}_\nu )
\end{equation}
is the periodic crystal potential. For simplicity, we consider a
non-degenerate $s$-type band (the case of degenerate bands, originating from
the atomic $d$- or $f$-functions, can be considered in a similar way\cite
{ii,iiz}). To pass to the second quantization representation we have to use
orthogonal wave functions. However, the atomic wave functions 
\begin{equation}
\varphi _\sigma ({\bf r},s)=\varphi ({\bf r})\chi _\sigma (s)
\end{equation}
($s$ is the spin coordinate) do not satisfy this condition at different
sites $\nu $. We apply the orthogonalization procedure developed by
Bogoliubov\cite{Bog}, which enables one to construct the perturbation theory
in the overlap of atomic wavefunctions belonging to different lattice sites.
To lowest order in this parameter the orthogonalized functions read 
\begin{equation}
\psi _\nu ({\bf r})=\varphi _\nu ({\bf r})-\frac 12\gamma \sum_{\nu ^{\prime
}\neq \nu }\varphi _{\nu ^{\prime }}({\bf r})  \label{psi}
\end{equation}
where the sum goes over the nearest neighbors, 
\begin{equation}
\gamma =\gamma _{\nu \nu ^{\prime }}=\int d{\bf r}\varphi _{\nu ^{\prime
}}^{*}({\bf r})\varphi _\nu ({\bf r})  \label{gam}
\end{equation}
is the non-orthogonality integral. Substituting (\ref{psi}) into (\ref{HV})
and using the orthogonality of the spin wave functions $\chi _\sigma (s)$ we
obtain the ME Hamiltonian of the ``polar model''\cite{sv,Bog,vk} 
\begin{eqnarray}
{\cal H} &=&\varepsilon \sum_{\nu \sigma }a_{\nu \sigma }^{\dagger }a_{\nu
\sigma }+\sum_{\nu _1\neq \nu _2,\sigma }t_{\nu _1\nu _2}a_{\nu _1\sigma
}^{\dagger }a_{\nu _2\sigma }  \label{H} \\
&&\ +\frac 12\sum_{\nu _i\sigma _1\sigma _2}I_{\nu _1\nu _2\nu _3\nu
_4}a_{\nu _1\sigma _1}^{\dagger }a_{\nu _2\sigma _2}^{\dagger }a_{\nu
_4\sigma _2}a_{\nu _3\sigma _1}  \nonumber
\end{eqnarray}
where 
\begin{equation}
\varepsilon =\int d{\bf r}\phi _\nu ^{*}({\bf r})[-\frac{\hbar ^2}{2m}\Delta
+v({\bf r-R}_\nu )]\phi _\nu ({\bf r})  \label{eps}
\end{equation}
is the one-electron level in the central potential of a given site $v({\bf r}%
)$, 
\begin{eqnarray}
t_{\nu _1\nu _2} &=&\int d{\bf r}\psi _{\nu _1}^{*}({\bf r})\left( -\frac{%
\hbar ^2}{2m}\Delta +V({\bf r})\right) \psi _{\nu _2}({\bf r})  \label{tt} \\
\ &\simeq &\int d{\bf r}\phi _{\nu _1}^{*}({\bf r})v({\bf r-R}_{\nu _1})\phi
_{\nu _2}({\bf r})  \nonumber
\end{eqnarray}
are the transfer matrix elements between the sites $\nu _1$ and $\nu _2$, 
\begin{equation}
I_{\nu _1\nu _2\nu _3\nu _4}=\int d{\bf r}d{\bf r}^{\prime }\psi _{\nu
_1}^{*}({\bf r})\psi _{\nu _2}^{*}({\bf r}^{\prime })\frac{e^2}{|{\bf r}-%
{\bf r}^{\prime }|}\psi _{\nu _3}^{}({\bf r})\psi _{\nu _4}^{}({\bf r}%
^{\prime })  \label{ipsi}
\end{equation}
are the matrix elements of interelectron Coulomb repulsion. We have
neglected in (\ref{eps}) and (\ref{tt}) the influence of potentials of the
sites $\nu ^{\prime }\neq \nu $ and $\nu ^{\prime }\neq \nu _1,\nu _2$
respectively, since the corresponding terms contain extra factors $v({\bf r-R%
}_{\nu ^{\prime }})\psi _\nu ({\bf r})$ with $\nu ^{\prime }\neq \nu $ which
are small due to the decrease of the potential $v({\bf r})$ with increasing $%
r{\bf \ }$(in other words, the crystal potential is small in the region
between the lattice sites). We have also dropped in (\ref{tt}) small
integrals which contain $v({\bf r-R}_{\nu ^{\prime }})|\psi _\nu ({\bf r}%
)|^2.$

Further we retain in (\ref{H}) only one- and two-site terms and pass to the
representation of the Hubbards operators $X\nu (\lambda ^{\prime },\lambda )$%
\cite{Hub4} which transform the state $|\lambda \rangle $ ($\lambda
=0,\sigma ,2$) at the site $\nu $ into $|\lambda ^{\prime }\rangle ,$%
\begin{equation}
a_{\nu \sigma }^{\dagger }=X_\nu (\sigma ,0)+\sigma X_\nu (2,-\sigma )
\end{equation}
Then we derive 
\begin{eqnarray}
{\cal H} &=&\varepsilon \sum_{\nu \sigma }X_\nu (\sigma ,\sigma )+U\sum_\nu
X_\nu (2,2)  \nonumber \\
&&\ \ +\sum_{\nu _1\neq \nu _2,\sigma }\{t_{\nu _1\nu _2}^{(00)}X_{\nu
_1}(\sigma ,0)X_{\nu _2}(0,\sigma )+t_{\nu _1\nu _2}^{(22)}X_{\nu
_1}(2,\sigma )X_{\nu _2}(\sigma ,2)  \nonumber \\
&&\ \ +\sigma [t_{\nu _1\nu _2}^{(02)}X_{\nu _1}(\sigma ,0)X_{\nu
_2}(-\sigma ,2)+X_{\nu _1}(2,-\sigma )X_{\nu _2}(0,\sigma )]\}  \nonumber \\
&&\ \ +\sum_{\nu _1\neq \nu _2}\{Q_{\nu _1\nu _2}n_{\nu _1}n_{\nu _2}-J_{\nu
_1\nu _2}[\frac 12+2({\bf S}_{\nu _1}{\bf S}_{\nu _2})]\}  \label{hh}
\end{eqnarray}
where 
\begin{equation}
n_\nu =\sum_\sigma n_{\nu \sigma }=\sum_\sigma a_{\nu \sigma }^{\dagger
}a_{\nu \sigma }=\sum_\sigma [X_\nu (\sigma ,\sigma )+X_\nu (2,2)],
\end{equation}
and 
\begin{equation}
U=I_{\nu \nu \nu \nu },\,\,Q_{\nu _1\nu _2}=I_{\nu _1\nu _2\nu _1\nu _2}
\end{equation}
are the Hubbard parameter and the Coulomb integral at different lattice
sites, 
\begin{eqnarray}
t_{\nu _1\nu _2}^{(00)} &=&t_{\nu _1\nu _2}  \label{t01} \\
t_{\nu _1\nu _2}^{(22)} &=&t_{\nu _1\nu _2}+2I_{\nu _1\nu _1\nu _2\nu _1}
\label{t02} \\
t_{\nu _1\nu _2}^{(02)} &=&t_{\nu _1\nu _2}^{(20)}=t_{\nu _1\nu _2}+I_{\nu
_1\nu _1\nu _2\nu _1}  \label{t03}
\end{eqnarray}
are the transfer integrals for empty states (holes) $|0\rangle $ and doubly
occupied states (doubles) $|2\rangle $, and the integral of the double-hole
pair creation.

Estimations of the parameters of the Hamiltonian (\ref{H}) with
non-orthogonality being neglected are presented in Ref.\cite{Hub}. In the
expressions for $\varepsilon ,U$ and $Q$ the non-orthogonality corrections
are small in the overlap and do not play a role. On the other hand, at
calculating other parameters in (\ref{hh}) we have to take into account the
second term in (\ref{psi}). We obtain for the integral of the ``direct''
exchange 
\begin{eqnarray}
\ \ J_{\nu _1\nu _2} &=&I_{\nu _1\nu _2\nu _2\nu _1}  \label{J} \\
\ &=&\widetilde{J}_{\nu _1\nu _2}-2\gamma _{\nu _1\nu _2}L_{\nu _1\nu _2}+%
\frac 12(U+Q_{\nu _1\nu _2})\gamma _{\nu _1\nu _2}^2  \nonumber
\end{eqnarray}
where 
\begin{equation}
\widetilde{J}_{\nu _1\nu _2}=\widetilde{I}_{\nu _1\nu _2\nu _2\nu
_1},\,L_{\nu _1\nu _2}=\widetilde{I}_{\nu _1\nu _1\nu _2\nu _1},
\end{equation}
the integrals $\widetilde{I}$ being calculated for the atomic functions $%
\varphi $. The expressions (\ref{t01})-(\ref{t03}) take the form 
\begin{eqnarray}
t_{\nu _1\nu _2}^{(00)} &=&t_{\nu _1\nu _2}  \label{t1} \\
\ t_{\nu _1\nu _2}^{(22)} &=&t_{\nu _1\nu _2}+2L_{\nu _1\nu _2}-(U+Q_{\nu
_1\nu _2})\gamma _{\nu _1\nu _2}  \label{t2} \\
t_{\nu _1\nu _2}^{(02)} &=&t_{\nu _1\nu _2}+L_{\nu _1\nu _2}-\frac 12%
(U\gamma _{\nu _1\nu _2}+Q_{\nu _1\nu _2})\gamma _{\nu _1\nu _2}  \label{t3}
\end{eqnarray}
All the terms in (\ref{J}), as well as in (\ref{t1})-(\ref{t3}), are of the
same (first) order in the overlap. Moreover, the quantity $U\gamma $ should
be larger than $L,Q,\widetilde{J}/\gamma $ and $|t|.$ Indeed, in the case of
narrow bands the interelectron repulsion (which determines $U,Q,\widetilde{J}
$ and $L$) and the crystal potential (which determines $\varepsilon $ and $t$%
) are of the same order of magnitude, although the crystal potential is
expected to be somewhat larger. At the same time, as follows from (\ref{tt}%
), (\ref{ipsi}), $t$ (or $Q,\widetilde{J}$ and $L$) contain another small
factor that was considered above: these integrals include the product of the
function $\phi _\nu ({\bf r})$ by the potential $v({\bf r-R}_{\nu ^{\prime
}})$ (or the electron repulsion) corresponding to other lattice site, which
``cuts'' a rather small region ${\bf r}\simeq {\bf R}_{\nu ^{\prime }}$.
Thus the ``on-site Coulomb'' (non-orthogonality) contributions should
dominate in (\ref{t2}) and (\ref{t3}).

It should be noted that in fact the transfer integrals (\ref{t2}) and (\ref
{t3}) are to be calculated with the use of many-electron wave functions (see
the review\cite{ii}) which are, generally speaking, not reduced to the
Slater determinants and not factorized into one-electron ones. For example,
the general Hartree-Fock approximation in the atom theory (see Ref.\cite{sob}%
) uses the radial ``one-electron'' wave functions which depend explicitly on
the ME atomic term $|\Gamma \rangle $. The transfer integrals are expressed
through the corresponding ME wave functions as\cite{ii} 
\begin{eqnarray}
&&t_{\nu _1\nu _2}(\Gamma \Gamma ^{\prime },\Gamma ^{\prime \prime }\Gamma
^{\prime \prime \prime }) 
\begin{tabular}{l}
=
\end{tabular}
\int \prod d\{{\bf r}_is_i\}\Psi _{\nu _1\Gamma }^{*}\Psi _{\nu _2\Gamma
^{\prime \prime }}^{*} \\
&&\ \ \ \ \ \times \sum_i\left( -\frac{\hbar ^2}{2m}\Delta _{{\bf r}_i}+V(%
{\bf r}_i)\right) \Psi _{\nu _1\Gamma ^{\prime }}^{}\Psi _{\nu _2\Gamma
^{\prime \prime \prime }}^{}  \nonumber
\end{eqnarray}
Therefore the integrals (\ref{t1})-(\ref{t3}) can be different even at
neglecting interatomic Coulomb interactions and non-orthogonality.

The electron spectrum of the model (\ref{hh}) in the simplest ``Hubbard-I''
approximation\cite{Hub} (which corresponds to a ``mean-field approximation''
in the electron hopping, the on-site Coulomb repulsion being taken into
account in the zero-order approximation) is given by\cite{ii} 
\begin{eqnarray}
E_{{\bf k}\sigma }^{(1,2)} &=&\varepsilon +\frac 12[t_{{\bf k}%
}^{(00)}\langle n_\sigma \rangle +t_{{\bf k}}^{(22)}(1-\langle n_{-\sigma
}\rangle )+U]  \label{h1} \\
&&\ \mp \frac 12\{[t_{{\bf k}}^{(00)}\langle n_\sigma \rangle -\ t_{{\bf k}%
}^{(22)}(1-\langle n_{-\sigma }\rangle )-U]^2  \nonumber \\
&&\ +4\left( t_{{\bf k}}^{(02)}\right) ^2\langle n_\sigma \rangle (1-\langle
n_{-\sigma }\rangle )\}^{1/2}  \nonumber
\end{eqnarray}
One can see that, unlike the standard consideration\cite{Hub}, the Hubbard
subbands (\ref{h1}) turn out to have quite different widths even in the
paramagnetic case for a nearly half-filled band ($\langle n_{+}\rangle
=\langle n_{-}\rangle \simeq 1/2$). In particular, in the large-$U$ limit we
have 
\begin{eqnarray}
E_{{\bf k}\sigma }^{(1)} &=&\varepsilon +(1-\langle n_{-\sigma }\rangle )t_{%
{\bf k}}^{(00)},\,\, \\
\,E_{{\bf k}\sigma }^{(2)} &=&\varepsilon +t_{{\bf k}}^{(22)}\langle
n_\sigma \rangle +U  \nonumber
\end{eqnarray}
so that, according to (\ref{t1}),(\ref{t2}), the bare hopping integral $t$
determines the bandwidth of ``holes'', and the bandwidth of ``doubles'' is
mainly (in the above-discussed sense) determined by the intrasite Coulomb
interaction and non-orthogonality integral (\ref{gam}). Thus an appreciable
asymmetry of the cases $N_e<N$ and $N_e>N$ can occur, the bandwidth in the
case of hole conductivity being considerably smaller than in the electron
(double) case. This circumstance may be important, e.g., for copper-oxide
high-$T_c$ superconductors. Of course, more advanced approximations than (%
\ref{h1}) should be used in the antiferromagnetic state, especially in the
two-dimensional case. In the latter situation, the low-energy electron
spectrum is determined by the scale $J$ rather than $t$ (see Ref.\cite{Lee}).

Generally speaking, we have to take into account in (\ref{hh}) also
three-site ``operator'' Coulomb contributions to the transfer integrals, 
\begin{equation}
\delta t_{\nu _1\nu _2}^{(\lambda \mu )}=\sum_{\nu ^{\prime }\neq \nu _1,\nu
_2}I_{\nu _1\nu ^{\prime }\nu _2\nu ^{\prime }}n_{\nu ^{\prime }}
\label{dt3}
\end{equation}
The quantity $\widetilde{I}_{\nu _1\nu ^{\prime }\nu _2\nu ^{\prime }}$ is
small in comparison with $L_{\nu _1\nu _2}=\widetilde{I}_{\nu _1\nu _1\nu
_2\nu _1}$ due to the decrease of the Coulomb interaction with distance. A
peculiar situation occurs in the case of a ``frustrated'' lattice where
equilateral triangles of nearest neighbors are present (e.g., the
triangular, fcc and hcp lattices), so that the site $\nu ^{\prime }$ can be
the nearest neighbor for both the sites $\nu _1$ and $\nu _2.$ Then
substituting (\ref{ipsi}) into (\ref{dt3}) yields the non-orthogonality
correction of order of $U\gamma .$ Such corrections are important for the
calculation of electron spectrum (e.g., in the ``Hubbard-I'' approximation)
and yield a suppression of the above-considered asymmetry of the hole and
double subbands. In such situations, additional three-site ``exchange''
terms also arise\cite{iiz}, so that the non-orthogonality problem\cite{mat}
is more difficult.

The kinetic exchange (Anderson's superexchange) interaction occurs in the
second order in $t_{\nu _1\nu _2}^{(02)}$. Performing the canonical
transformation which excludes the double-hole pair creation and annihilation
terms from the Hamiltonian (\ref{hh}) we derive 
\begin{equation}
{\cal H}_{}^{(2)}=\sum_{\nu _1\nu _2}\frac{\left( t_{\nu _1\nu
_2}^{(02)}\right) ^2}U\{4({\bf S}_{\nu _1}{\bf S}_{\nu _2})-1\}  \label{ex}
\end{equation}
As follows from (\ref{t3}), the nominator in (\ref{ex}) is determined not
only by the bare hopping, but also by the Coulomb interaction. Formally, the
kinetic exchange interaction survives even in the limit $U\rightarrow \infty 
$ owing to the non-orthogonality contributions. Combining (\ref{ex}) and (%
\ref{J}) we obtain the expression for the total effective exchange parameter
in the case of a half-filled band ($c=0$) 
\begin{eqnarray}
J_{eff} &=&J-2\left( t^{(02)}\right) ^2/U  \nonumber \\
\ &=&\widetilde{J}-\gamma ^2Q+2\gamma t-2(t+L)^2/U  \label{Jef}
\end{eqnarray}
Note that the terms of the order of $\gamma ^2U$ are canceled in $J_{eff}$.
First three terms in (\ref{Jef}) coincide with the corresponding result for
the two-site problem (hydrogen molecule) \cite{mat} and yield an
antiferromagnetic exchange interaction. As mentioned above, the crystal
potential should be somewhat larger than the Coulomb interaction, and in the
large-$U$ limit the main contribution to $J_{eff}$ reads 
\begin{equation}
J_{eff}\simeq -2\gamma |t|
\end{equation}
where $t$ is assumed to be negative. We see that in the case $N_e<N$ the
ratio of $J_{eff}$ to bandwidth is proportional to the overlap parameter
(rather than to $|t|/U\ll \gamma $ as in the standard consideration). High
values of the N\'eel temperature, which are typical for the layered
copper-oxide systems, may be related to this fact.

By analogy with the consideration of Ref.\cite{Nag} we have the criterion of
ferromagnetism 
\begin{equation}
2\alpha c|t^{(\lambda \lambda )}|>-J_{eff}  \label{new}
\end{equation}
where $\lambda =0$ for $N_e<N$ and $\lambda =2$ for $N_e>N.$ Under the
assumption $\gamma \gg |t|/U\gg \gamma ^2$ the criterion (\ref{new}) takes
the form 
\begin{equation}
\alpha c>\left\{ 
\begin{tabular}{ll}
$-J_{eff}/(2|t|)\simeq \gamma ,$ & $N_e<N$ \\ 
$-J_{eff}/(\gamma U)\simeq 2|t|/U,$ & $N_e>N$%
\end{tabular}
\right.  \label{new1}
\end{equation}
which is quite different from (\ref{Nf}) for $N_e<N$. In the case $N_e>N$
the result (\ref{new1}) is formally similar to (\ref{Nf}), but has a
different origin.

We see that a strong indirect antiferromagnetic interaction occurs in narrow
energy bands, the ferromagnetic exchange owing to the motion of current
carriers being more strongly suppressed in the case of ``hole'' conductivity
($N_e<N$) than in the ``electron'' case. The situation may change
considerably in the case of degenerate energy bands where the indirect
interaction can be ferromagnetic owing to the intraatomic Hund exchange\cite
{Khom,ii}. Note that many-band effects are often assumed to be essential
also for the usual itinerant magnetism (see, e.g., Ref.\cite{mat}).

From the experimental point of view, the narrow-band ferromagnetism is not a
too wide-spread phenomenon. It takes place, e.g., in the systems Fe$_{1-x}$Co%
$_x$S$_2$\cite{pir}, CoS$_2,$ CrO$_2$\cite{Mott}. However, the degeneracy of
conduction band plays an important role in the electron structure of these
systems. At the same time, ferromagnetism is not observed in the
copper-oxide systems. The above-discussed modifications (in comparison with
the original Hubbard's treatment\cite{Hub}) in formulation of the simplest $s
$-band model and similar considerations of more realistic models (see  Refs.%
\cite{tJ,Bel}) may be useful for explaining this fact.

The research described was supported in part by the Grant No.96-02-1600 from
the Russian Basic Research Foundation.

\end{document}